\documentstyle[12pt,blois]{article}

\begin{document}

\input psfig.tex

\heading{THE AGE OF EARLY--TYPE GALAXIES\\ 
         IN CLUSTERS\\
         AN OBSERVATIONAL PERSPECTIVE}

\author{S. Andreon} {Osservatorio Astronomico di Capodimonte, Napoli, Italy.}

\begin{bloisabstract}
Contradictory results on the evolution of
lenticular galaxies (S0s) at intermediate redshift have been 
recently published. A careful analysis of the used methodology shows that
in order to gain insights on the evolution of the morphological types it is
necessary:  (1) to use quantitative morphological types, (2) to propagate
morphological errors in the quantity used to measure the evolution of
the S0s and (3) to adopt 
the same morphological classes at high as at low redshift, testing 
the correspondence with standard types. 
Some published analysis fail to
satisfy at least one of the above three conditions, 
thus undermining the claim for a young age of S0s of 
intermediate redshift clusters.
\end{bloisabstract}

\section{Introduction}

Two contradictory results on the evolution of lenticular (S0) galaxies 
are found: either S0s are old ($z_{formation}>2$) and
are evolving passively (\cite{BLE92}, \cite{ESD}, \cite{ADH97},
\cite{SED}, \cite{L98}) or most of them formed at $z<0.5$, as implied by
the claimed deficit of S0s in intermediate redshift ($z\sim0.5$) clusters
(\cite{S97}, \cite{D97}). Furthermore, \cite{vD} found that at large 
clustercentric radii ($R>0.7
h^{-1}_{50}$ Mpc) of the intermediate redshift cluster Cl 1358+62 S0s are
heterogeneous in color and have therefore experienced 
recently star formation. 

This situation is quite puzzling: intermediate redshift lenticulars are
red and with little scatter in color and are therefore thought to be as
old as ellipticals (\cite{ESD}, \cite{ADH97}). Furthermore, data and samples 
studied by the different teams largely overlap: \cite{ADH97} and
\cite{ESD} studied respectively
one and three clusters of the sample of 9
clusters studied by \cite{S97} \& \cite{D97}.  Are the differences
found by the various teams due to the different analysis of the data or
they reflect the possibility that
the evolutionary history of galaxies differs from cluster to cluster, as
may be suggested by the heterogeneous composition of the three intermediated
clusters studied by \cite{C97}?

All the quoted works share a common idea: take a sample of galaxies
in cluster, possibly at different redshift, split it
in (morphological) classes, measure a property of the class
(the morphological composition or the scatter in color of each class)
and derive from this measure an evolutionary path for the classes.
In this paper we focus our attention on the way galaxies are classified
and on the effect that classification errors have on the inferred galaxy
evolution. A more detailed discussion is
given in \cite{A98}. \cite{Brew} focus their
review on how a given scatter in color propagates on the formation age.

\begin{figure}
\centerline{%
\psfig{figure=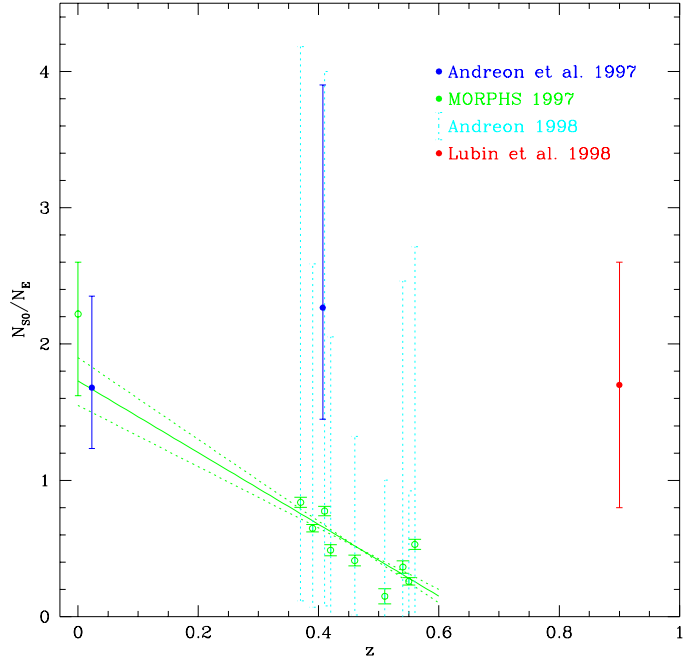,bbllx=70pt,bblly=200pt,bburx=550pt,bbury=670pt,width=9truecm}}
\caption{The S0 to E ratio as a function of z. The small
errorbars (slightly larger
than the symbol size) of the clusters at $z\sim0.5\pm0.1$ and at
small S0 to E ratio are the original
estimate (\cite{D97}). The large errorbars instead are computed
by the author taking
into account the morphological errors. The plotted lines
are the best fit to the intermediate redshift S0 to E ratios \cite{D97}.}
\end{figure}

\section{The effect of misclassification errors on the S0 to E ratio}

The spiral fraction in clusters increases with redshift, which implies
that the E+S0 fraction decreases. The real variation of the E and S0
fractions due to evolution is therefore easily masked by the change
of the S fraction.  The S0 to E ratio, instead, is less dependent on the
change of the S fraction and can be used as an indicator of a differential
evolution of Es and S0s.

Figure 1 shows the various determinations of the S0 to E ratio.
In the local universe, various determinations of the S0 to E ratio give
similar values: $2.2^{+0.2}_{-0.3}$ \cite{D97}, $1.7^{+0.35}_{-0.25}$
\cite{A96}, the latter being based on only one cluster, whereas the
former is the average of the composition of 
$\sim50$ clusters (whose completeness in magnitude
is undefined). For one cluster at intermediate ($z\sim0.4$) redshift and
for another at high redshift ($z\sim0.9$),
\cite{ADH97} and \cite{L98} found a ratio of
$2.2^{+1.6}_{-0.7}$  and $1.75\pm0.9$,
respectively. Instead
\cite{D97} found for 9 clusters at intermediate redshift a ratio 
of $0.5\pm0.05$. 

Reference \cite{D97} assumes that classification errors
have a negligible impact on the S0 to E ratio.
Given the errors, {\it each} S0 to E ratio at $z\sim0.4$ measured by \cite{D97}
is incompatible at $1-2\sigma$
with the values measured by other teams or by themselves at 
zero redshift;  \cite{D97} therefore claim the existence of a deficit of S0s in
intermediate redshift from their own measures.

Morphological types, as all physical quantities, are
subject to errors. 
Usually, 20 \% of the galaxies have discrepant morphological types when
classified by independent morphologists (\cite{AD97}), and this holds 
true also
for the morphological estimates made by \cite{D97} (cfr. \cite{S97}). 

How large is the impact of morphological errors on the S0 to E ratio? 

If the number of misclassified galaxies of a given
class, say E, is balanced by the number of the 
misclassified galaxies of the other
classes, say S and S0, 
morphological errors have negligible effects.
However, the amplitude of the error depends on the cluster
morphological compositions, since
in clusters with different compositions
a different number of galaxies enter and exit from each class because
the probability of a correct (or uncorrect) classification
is independent on the morphological composition of the cluster.
For this reason, the impact of morphological errors is not always 
negligible.

Actually, the impact may be
very large, say larger than 100 \%, when the S0 to E ratio is small
(as in the estimate of \cite{D97} at intermediate redshift),
or when the morphological errors are systematic, 
for example if more galaxies of
type X are misclassified as of type Y than vice versa.
At first sight, the assumption that the errors are largely systematic
seems excessive, nevertheless we note that this is what happens in the
cluster in common between \cite{ADH97} and \cite{D97}:  
all S0s in \cite{ADH97} are classed E or S by \cite{D97}
and no-one E or S is misclassified as S0. 

Taking into account morphological errors, 
(\cite{D97})'s errorbars grow
from $\sim0.05$ to $\sim2$ units, making all the estimates of the S0 to E ratio equal
within less than $1\sigma$ (see Figure 1), thus undermining
the claimed deficit of S0s at intermediate redshift.

\medskip
We would like to underline that
the claim of a deficit of S0s in intermediate redshift
clusters is based on a subjective measure: the estimate of the
morphological type by visual inspection of the galaxy images.
It is unsatisfactory that the evolution of a class of object is ultimately 
based
on a qualitative measurement (the morphological type). It would be
much more reliable if the possible existence of such a deficit
could be based on a
quantitative estimate of the morphological type. It should
be stressed, however, that a small sample have such a
quantitative morphological estimate 
and does not present any evidence for a deficit of S0s (\cite{ADH97}).

\setbox11=\hbox{$\vcenter{\psfig{figure=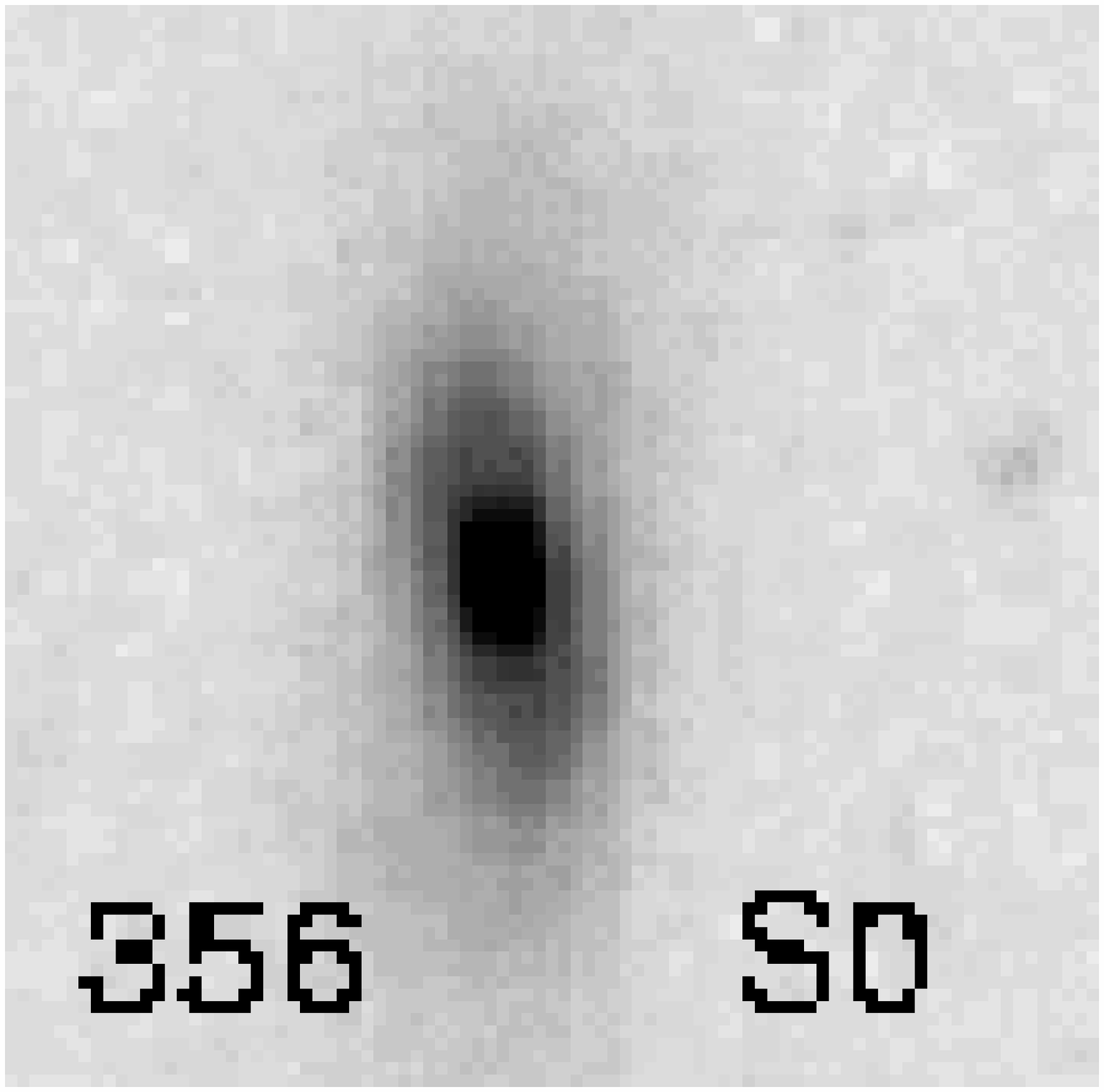,width=3truecm}}$}
\setbox12=\hbox{$\vcenter{\psfig{figure=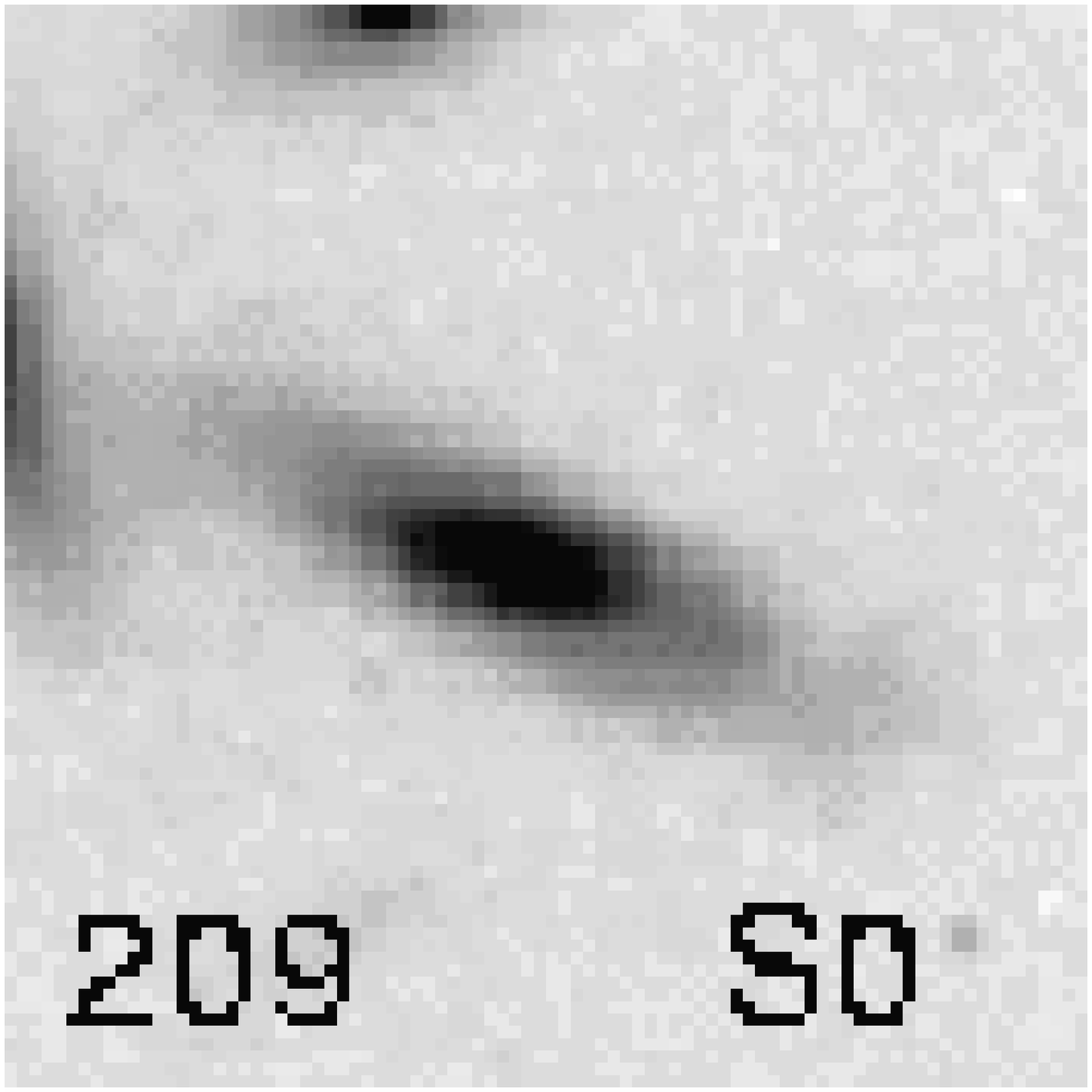,width=3truecm}}$}
\setbox13=\hbox{$\vcenter{\psfig{figure=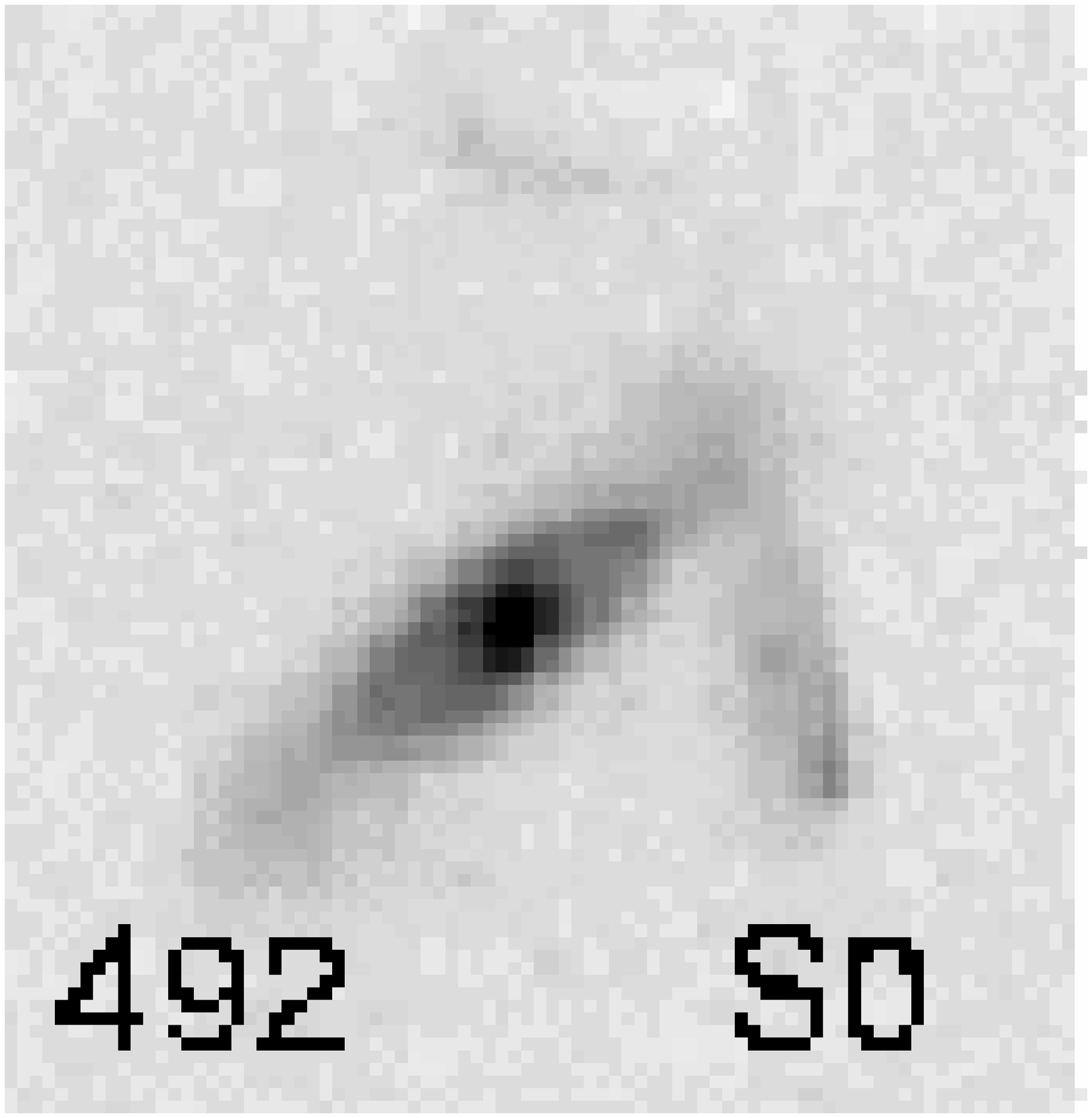,width=3truecm}}$}
\setbox14=\hbox{$\vcenter{\psfig{figure=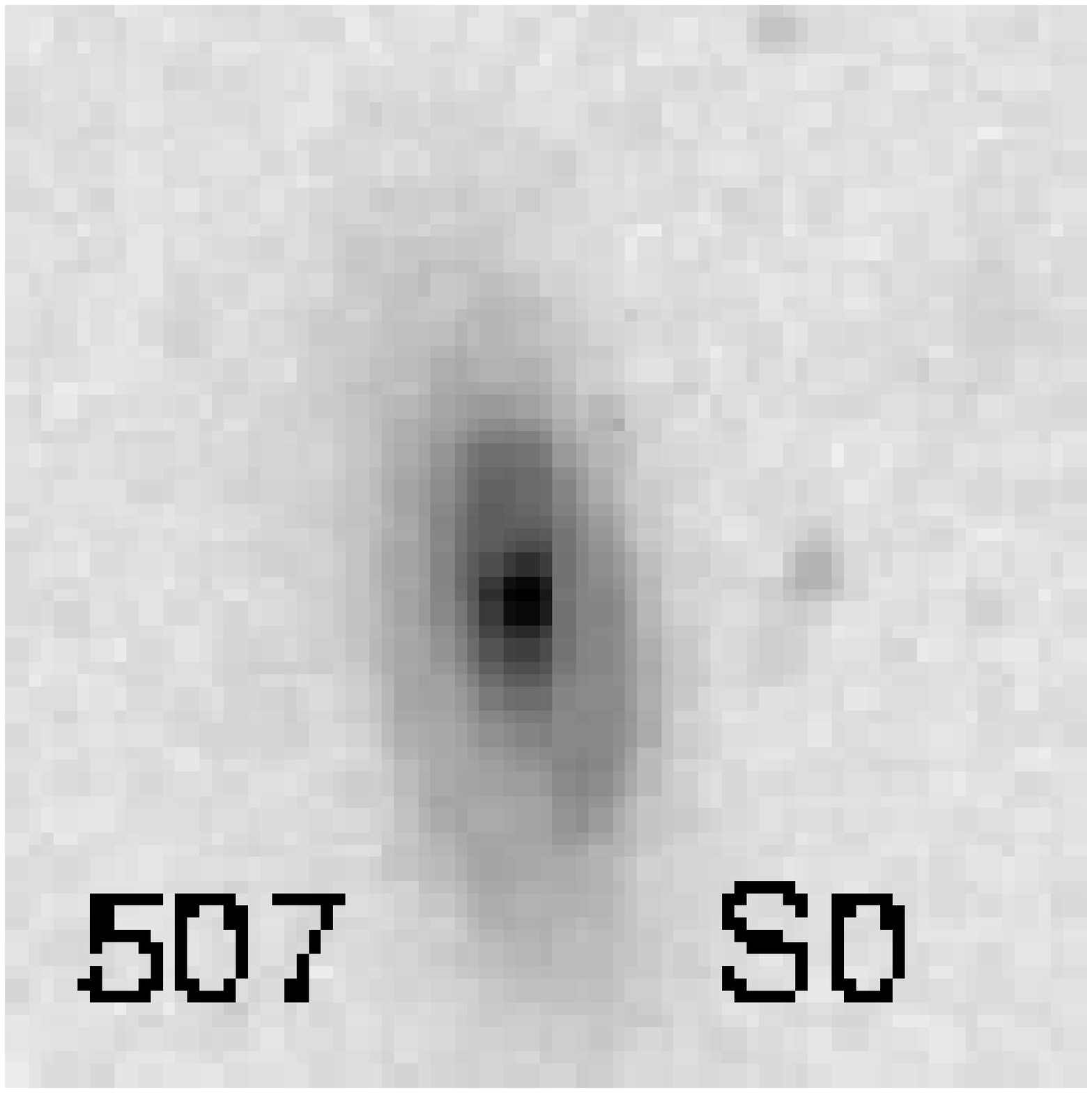,width=3truecm}}$}
\setbox15=\hbox{$\vcenter{\psfig{figure=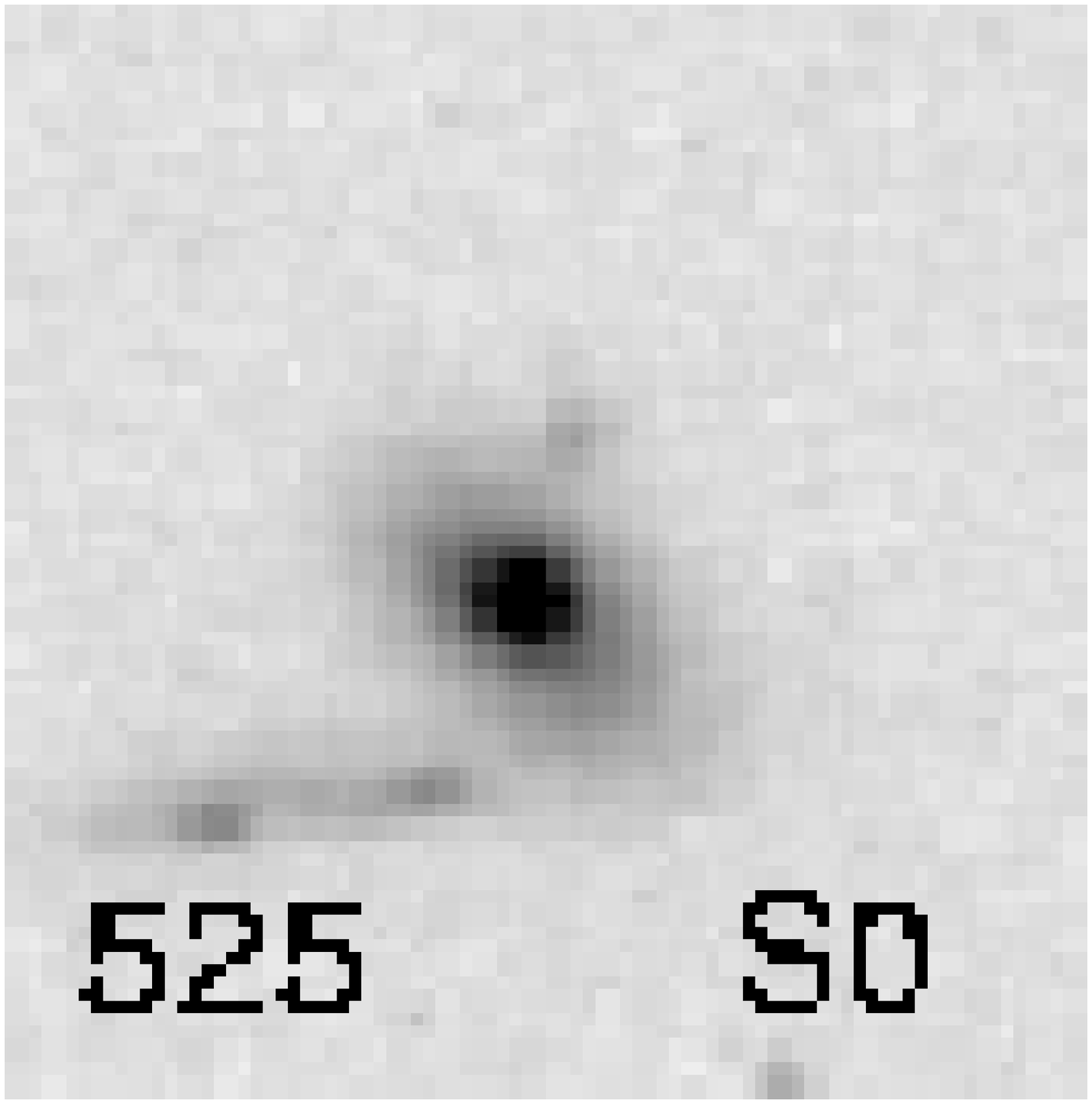,width=3truecm}}$}
\setbox9=\hbox{$\vcenter{\psfig{figure=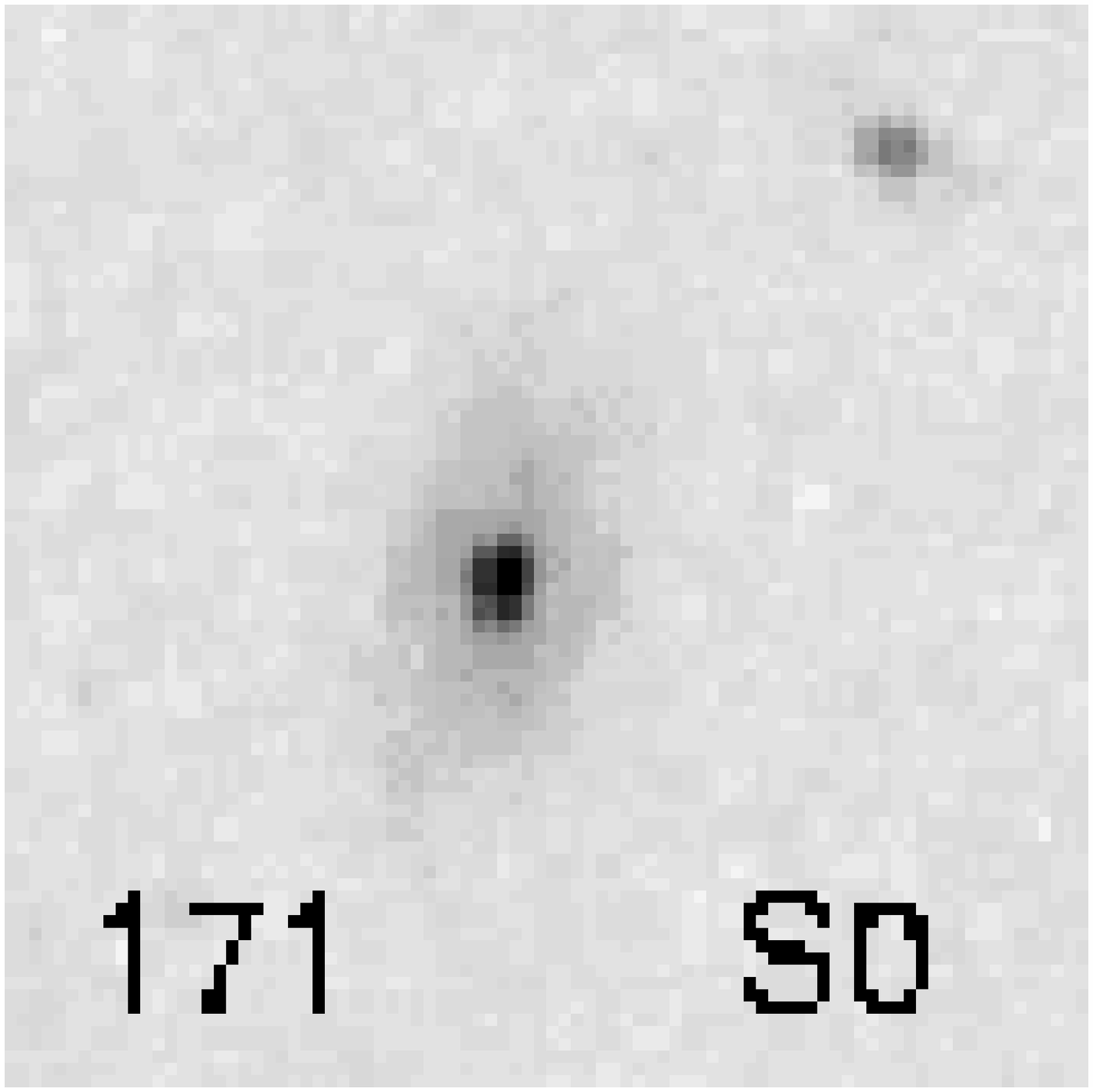,width=3truecm}}$}

\begin{figure}
\centerline{\box9 \quad \box12 \quad \box11}
\centerline{\box13 \quad \box14 \quad \box15}
\caption{Postage stamp of some galaxies classified as S0 by \cite{vD},
but whose usual classification is S.}
\end{figure}

\section{The heterogeneity of S0s at large clustercentric radii}

Recently, \cite{vD} found that at large clustercentric radii
($R>0.7 h^{-1}_{50}$ Mpc) in the intermediate redshift cluster Cl 1358+62,
S0s are heterogeneous in color and therefore experienced star formation
until very recently; a conclusions which is opposite to those claimed in
previous
works focused on the central regions of other intermediate redshift
clusters (\cite{ESD}, \cite{ADH97}). They suggest that S0s evolve primarily
in the transition region between the cluster and the field, giving support
to \cite{D97}'s \& \cite{S97}'s finding that S0s are still forming at
intermediate redshift.
However, \cite{vD} choose to classify as S starforming galaxies only,
leaving spirals with faint or smooth spiral structures in the S0 class (see
their Section 2.3.1). This is confirmed by the inspection of their black
and white prints presented in their Figure 3: at least 25 \% of all
galaxies that \cite{vD} classify as S0s are S. Some notable examples are
reproduced here in Figure 2. All these galaxies show smooth spiral arms
(or irregular isophotes) and look as Cl 0939+47 or Coma
spirals.  Therefore, \cite{vD} use a classification scheme different
from the one of all the other workers.  Their conclusion on the
evolutive nature of S0s is therefore relative to their S0 class, and not
to the Hubble S0 class. 

\section{Conclusion}

The analysis of the work done so far shows that in order to gain insights
on the evolution of galaxies at intermediate redshift we need:

-- to use quantitative morphological types: the conclusions drawn from
splitting galaxies in classes are too important
to be based on qualitative measures only;

-- to take into account the effect of morphological errors on the quantity
used to measure the evolution of the considered class;

-- to adopt the same morphological
classes at high and at low redshift, testing the ability to
reproduce the morphological type assigned by famous and recognized
morphologists. Hubble types (E, S0, S) correspond to well defined
morphologies and should not be used for other types of objects.

The two works presenting evidences for different evolutionary
paths for S0 and E fail to satisfy at least one of these three conditions,
thus undermining the claim for a young age of S0s of 
intermediate redshift clusters.

\acknowledgements{We thank the organizer of the ``Xth Rencontres de
Blois", for the financial support given for attending the conference.
We thank G. Longo for an attentive lecture of this paper.}


%
\vfill
\end{document}